\pdfoutput=1
\documentclass[11pt]{article}
\usepackage{multirow}
\usepackage{hyperref}
\usepackage{makecell}
\usepackage[numbers]{natbib}

\usepackage{graphicx}  
\usepackage{float}    

\usepackage{amsmath}
\usepackage[margin=1.2in, top=1.2in]{geometry}
\usepackage{caption}
\usepackage{times}
\usepackage{latexsym}
\usepackage[T1]{fontenc}
\usepackage{booktabs}

\usepackage[utf8]{inputenc}
\usepackage{microtype}
\usepackage{inconsolata}
\usepackage{graphicx}

\usepackage{comment}
\usepackage{parskip}  

\title{\textbf{Embracing AI in Education: Understanding the Surge in Large Language Model Use by Secondary Students}}
\date{}

\author{
  \textbf{Tiffany Zhu\textsuperscript{1}},
  \textbf{Kexun Zhang\textsuperscript{2}},
  \textbf{William Yang Wang\textsuperscript{3}}, \\
  \textsuperscript{1}The Harker School,
  \textsuperscript{2}Carnegie Mellon University,
  \textsuperscript{3}University of California, Santa Barbara
  \\
  \small{
    \textbf{Correspondence:} \href{mailto:26tiffanyz@gmail.com}{26tiffanyz@gmail.com}
  }
}

\begin{document}
\maketitle

\begin{abstract}

The impressive essay writing and problem-solving capabilities of large language models (LLMs) like OpenAI's ChatGPT have opened up new avenues in education. Our goal is to gain insights into the widespread use of LLMs among secondary students to inform their future development. Despite school restrictions, our survey of over 300 middle and high school students revealed that a remarkable 70\% of students have utilized LLMs, higher than the usage percentage among young adults, and this percentage remains consistent across 7th to 12th grade. Students also reported using LLMs for multiple subjects, including language arts, history, and math assignments, but expressed mixed thoughts on their effectiveness due to occasional hallucinations in historical contexts and incorrect answers for lack of rigorous reasoning. The survey feedback called for LLMs better adapted for students, and also raised questions to developers and educators on how to help students from underserved communities leverage LLMs' capabilities for equal access to advanced education resources. We propose a few ideas to address such issues, including subject-specific models, personalized learning, and AI classrooms.

\end{abstract} 

\section{Introduction}

The remarkable generative capabilities of InstructGPT \cite{instructgpt}, introduced over two years ago, has led to widespread usage of large language models (LLMs). Since then, the capabilities of LLMs like OpenAI's ChatGPT \cite{chatgpt}, Google Gemini \cite{geminiteam2024gemini} and Meta Llama \cite{llama} have advanced significantly. According to an OpenAI report \cite{report}, GPT-4 achieves above the 90th percentile on exams like AP Biology, AP Calculus BC, and SAT math. Consequently, many middle and high school students have begun using LLMs for schoolwork assistance.

This growth in capabilities of artificial intelligence (AI) raises concerns about its impact on students’ learning experiences \cite{Kasneci}. We have observed schools implement policies restricting students from using LLMs due to concerns for cheating as well as inaccurate answers. Even OpenAI cautions secondary students against using ChatGPT, and specifically prohibits usage by children under the age of 13 \cite{policies}. While OpenAI is developing ChatGPT Edu for educational purposes\cite{chatgpt-edu}, its audience is only university students. Consequently, the secondary school age group lacks support and influence over the services and developments of LLMs \cite{instructgpt}. 

 Prohibited or not, access to chatgpt.com is very easy with no registration requirement. Such openness makes it challenging to restrict LLM usage by students. Instead of imposing restrictions, we believe the best approach is to understand how students utilize LLMs and develop curriculum and adapt LLMs to accommodate them. While some research \cite{Baidoo-anu,Adeshola} has been conducted on personalized learning with LLMs, large-scale comprehensive surveys understanding the trend of student usage of LLMs are lacking. Our study aims to fill this gap, as shown in Table \ref{tab:survey_comparison}.

\begin{table}[htbp]
  \centering
  \caption{Our study has more participants than past studies that investigated LLM Usage}
    \vspace{-0.4em}  
  \begin{tabular}{c c c c}
  \toprule
    \textbf{Study} &  \textbf{Demographic} &  \textbf{Participants} & \textbf{Purpose} \\ 
    \midrule
    Li et. al \cite{Li} & Chinese secondary schoolers & 76 & ChatGPT usage motives \\ 
    Belghith et. al \cite{Belghith} & Middle school students & 24 & AI interactions \\ 
    Rahman et. al \cite{Rahman} & College programming students & N/A & ChatGPT for coding \\ 
    This study & Middle and high schoolers & 306 & LLM trends \& potential \\ 
    \bottomrule
  \end{tabular}
  \label{tab:survey_comparison}
  \vspace{-0.5em} 
\end{table}

\begin{figure}[H]  
  \centering
  \includegraphics[width=6cm,height=3.4cm]{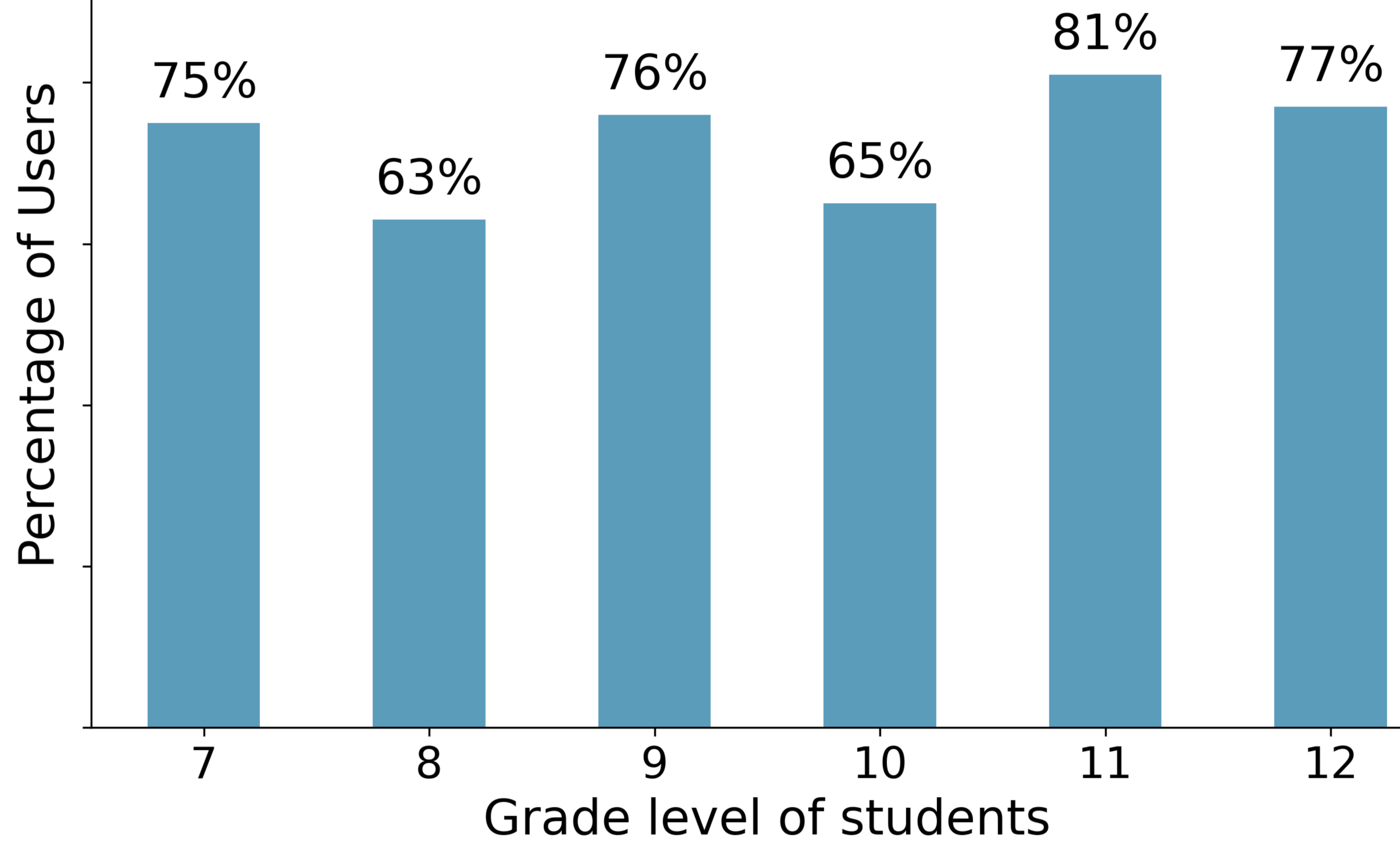}
  \vspace{-.5em}  
  \caption{LLM usage remained consistent among 7th-12th graders.}
  \label{fig:usage_by_grade}  
\end{figure}

A key question we sought to address was the prevalence of LLM usage among middle and high school students. Our survey confirmed our suspicion of its widespread usage,  with over 70\% of respondents reporting usage, higher than the 43\% usage among young adults as reported in a February poll \cite{Pew}. Surprisingly, we found the percentage of students using LLMs was quite consistent across grades 7 to 12, as shown in Figure \ref{fig:usage_by_grade}. We had anticipated lower usage among students in lower grades. 

  \vspace{-.5em}  

In the subsequent sections of this paper, we will present the details of our survey methodology and the results obtained for a range of other questions, including:
  \vspace{-0.4em} 
\begin{itemize}
    \item What subjects do students seek assistance from LLMs and how satisfied are they?
    \item How stringent are school policies regarding LLM usage and how do students regard them?
    \item Are there demographic differences among students that affect LLM usage?
\end{itemize}
    
To address the survey feedback, we will present proposals on how to further improve LLMs for education and embrace LLMs in classrooms to enhance learning for students from different backgrounds.

In summary, our contributions are several fold. First, with over 300 respondents across the United States, we obtained a collective view of LLM usage from a large, diverse group of secondary students. Second, we drew insights on critical LLM usage questions that are informative for educators, policymakers, and developers to embrace AI in education. Third, we provided proposals to integrate LLMs into education more effectively and responsibly.

\section{Data Collection}
\label{gen_inst}

\textbf{Target group} Our target group consisted of middle and high school students throughout the United States and encapsulated a diverse population. 

\textbf{Collection methods} The survey form took just a few minutes to complete, which encouraged respondents to take the survey without feeling lethargic and answering randomly. We used Centiment \cite{centiment}, a survey platform that allowed for connections with respondents throughout the United States, who were compensated over 80 cents for completing the short survey. We also created a Google Form survey, which we shared by visiting local tutoring programs and posting on social media.

The following information was provided to the survey takers: \textit{"Results of this survey are completely confidential and will be used for data analysis."} All responses are available in a public \href{https://github.com/26tiffanyz/Secondary-Student-LLM-Usage}{GitHub repository} with the respondents' personally identifiable information removed. 

\begin{figure}[H]  
  \centering
  \includegraphics[width=6cm,height=3.5cm]{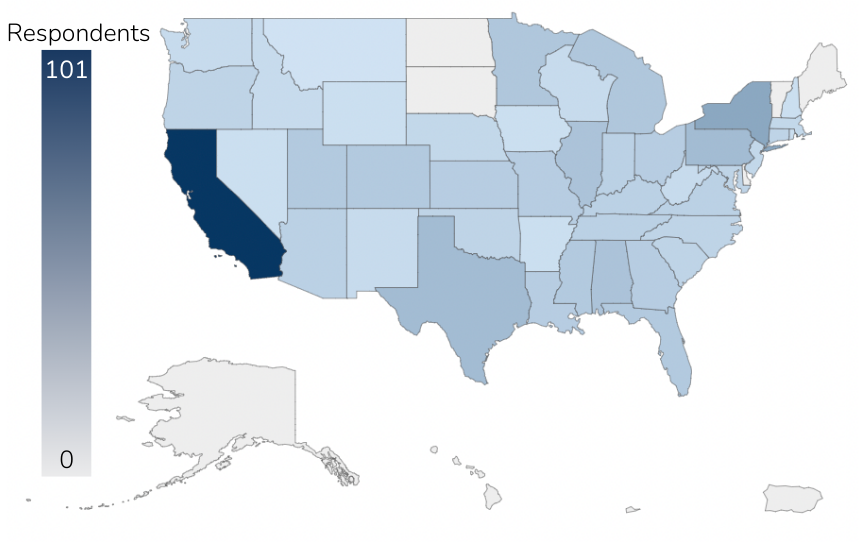} 
  \vspace{-.8em} 
  \caption{We received responses from students in 43 states.}
  \label{fig:students_by_state}  
\end{figure}

\textbf{Response demographics} Responses included students ranging from sixth to twelfth grade (Table \ref{tab:students_by_grade}). California had the highest number of responses, as the in-person surveys were conducted in this state, and we received feedback from a total of 43 states (Figure \ref{fig:students_by_state}). The respondents also came from various educational backgrounds, including private, public, charter, and home schooled. Lastly, the demographic encompassed students who consistently had access and students who seldom had access to technology for personal purposes. 

\begin{table}[htbp]
    \vspace{-.2em} 
  \centering
  \caption{Number of responses from students in grades 6-12}
  \vspace{-0.4em}  
  \begin{tabular}{c c c c c c c c}
    \toprule
    \textbf{Grade} & {6th} & {7th} & {8th} & {9th} & {10th} & {11th} & {12th}\\ 
    \midrule
    \textbf{Respondents} & 3 & 17 & 56 & 99 & 52 & 48 & 31\\ 
    \bottomrule
  \end{tabular}
  \label{tab:students_by_grade}
  \vspace{-0.5em} 
\end{table}

\section{Data Analysis}

\begin{figure}[H]
    \centering
    \includegraphics[width=6cm,height=5.5cm]{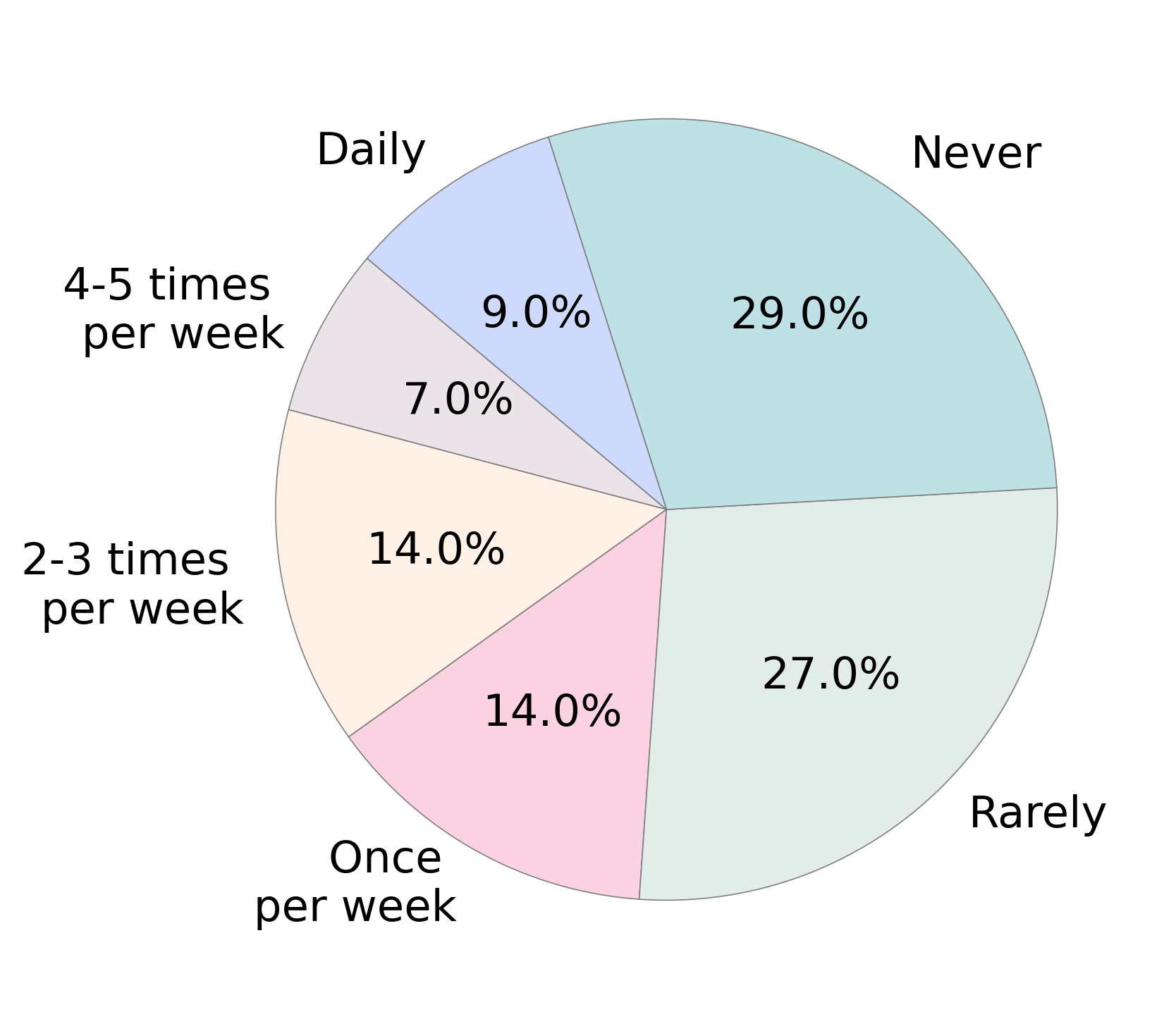}
    \vspace{-.9em}
    \caption{Over 70\% of students have used LLMs.}
    \label{fig:usage_per_week}
\end{figure}

Figure \ref{fig:usage_per_week} shows the usage frequency per week of LLMs, including ChatGPT, LLaMA, Gemini and Bing Copilot. Our survey shows that 71\% of the correspondents have used LLMs at least once, while 9\% used them daily. Since our survey shows 84\% of the LLM users use ChatGPT, this translates to 60\% of ChatGPT users among all the respondents. In comparison, 43\% of young adults of age 18-29 have  used ChatGPT\cite{Pew}. Our survey clearly shows popularity of LLMs among the secondary students and consistency across the grades, as reported in Figure \ref{fig:usage_by_grade}.

\begin{figure}[H]
    \centering
    \includegraphics[width=10cm]{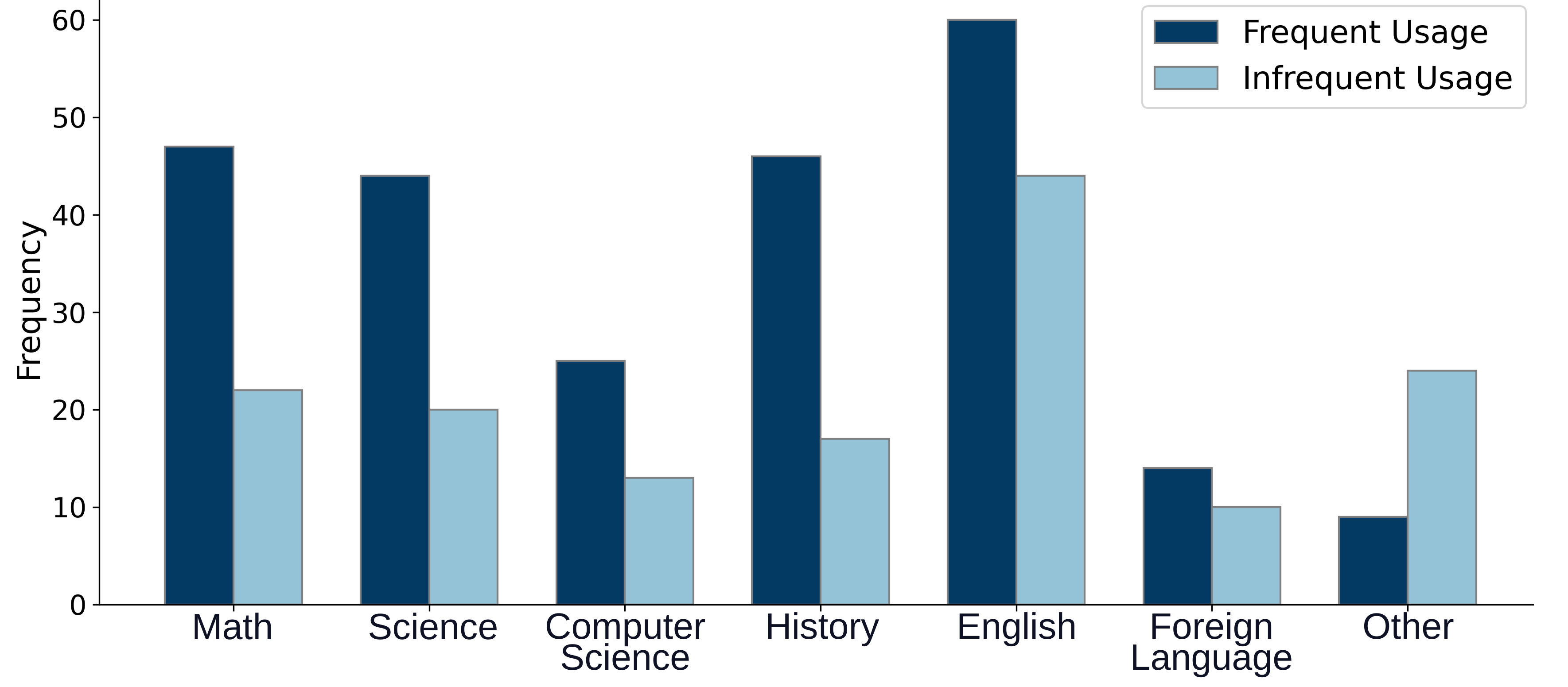}
    \vspace{-.9em}
    \caption{LLMs are used in all school subjects.}
    \label{fig:subject_usage}
\end{figure}

\textbf{LLMs are used in all school subjects.} 
Previous research reveals that LLMs exhibit inferior mathematical abilities compared to addressing open-ended requests in English \cite{Frieder}. Despite this, our survey shows 28\% students still used LLMs for the math subject. They also use LLMs in various other subjects, including foreign languages and history classes (Figure. \ref{fig:subject_usage}). According to our survey, 57\% of LLM users have reported finding them useful for writing text, while 43\% have found them useful for math and STEM questions. We also noticed that among students who use LLMs at least twice a week, the majority of them have tried LLMs for all class subjects. 

    \vspace{-.9em}

\textbf{Students share various thoughts regarding LLM impact.} Responses to an open-ended question regarding the impact of LLMs on students' academic performance were diverse, despite the common usage. The following are a few student responses with contrasting thoughts on LLMs' helpfulness.

\textit{“Yes, they have helped simplify complex topics.” \newline
“No, increased reliance on such tools may lower the level of other fundamental skills that you develop.” \newline
“Sometimes, as it may be inaccurate at certain points.” 
}

Some students, like the last, believe LLMs have inaccuracies when answering factual questions, which is supported by previous research that discovered ChatGPT hallucinates the most when questions are inputted consecutively \cite{Ahmad}. However, many other students fail to recognize the inaccuracies \cite{wang}.

\textbf{Most students desire LLM improvements for schoolwork.} When asked about potential improvements to LLMs, 3\% of individuals expressed no desire for any modifications. 51\% desired for more coherent responses, accurate answers and a great ability to address complex questions, indicating somewhat dissatisfaction with the current models. Only 20\% of respondents expressed interest in improving LLMs' understanding of emotions and better conversations, indicating students’ preference for using LLMs to answer questions rather than using LLMs as a conversational tool. 

\begin{figure}[H]
  \centering
  \includegraphics[width=5.3cm]{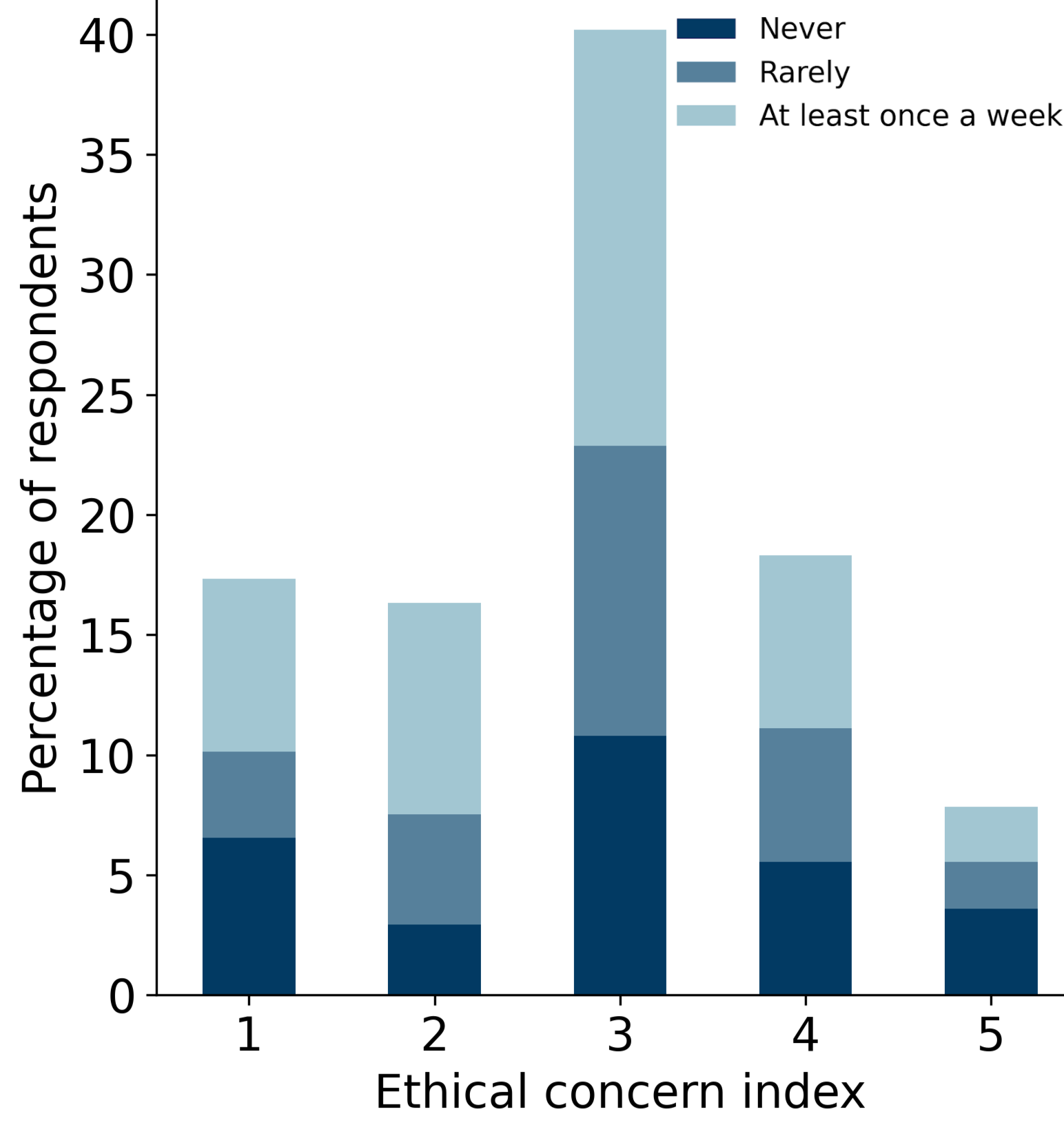}
  \vspace{-.6em}
  \caption{LLM usage persists despite ethical concerns.}
  \label{fig:usage_vs_ethical}
\end{figure}

\textbf{LLM usage is prevalent despite school policies.} Participants were asked to rate from 1 to 5 how strict their school policies are on getting help from LLMs, with a score of 5 meaning severe punishment if LLM usage was found in their schoolwork. The average rating was 3.48, signaling generally restrictive policies. 

We asked students to rate their ethical concerns on using LLMs, and got an average rating of 2.83 on a scale of 1-5, where a higher number meant a larger concern. The value is slightly less than what they perceived as severity of punishment rating, signaling usage outweighs ethical concerns, as shown by figure \ref{fig:usage_vs_ethical}. Interestingly, we found that 2\% of the students who rated themselves as a 5 on ethical concern reported using LLMs at least once a week, indicating that resisting using LLMs can be difficult. 

\textbf{More technologically advanced regions have more users.} According to the 2022 State Technology and Science Index, which ranks the states based on their technological economies, the five highest ranking states are Massachusetts, California, Colorado, Maryland, and Utah \cite{milken}. We discovered that 80.2\% of students in these five states and 64.3\% of students in the rest of the states had used LLMs. This difference of over 15\% shows a concerning gap between students with different tech savviness levels. However, it also points to a trend of increasing LLM usage among students since LLMs' continuing improvement and decreasing access barrier will attract more users from states with lower tech benchmarks .

We also discovered a small usage gap between private and public school students. Our survey shows 76.7\% of private school students reported LLM usage versus 71.3\% of public school students.  Since students attending private schools tend to be from wealthier families, this 5\% difference is likely caused by family income differences.

\textbf{Users who can pay for models have an advantage.} Out of the 17 participants who possess a paid version of the models, the outcomes regarding its usefulness are predominantly in favor of a positive response. Of these 17, all except one reported using LLMs at least 2-3 times a week, significantly more frequent than students using free models. As companies release higher-quality paid versions, students who have paid models will gain a greater advantage. This further proves the existence of an LLM usage gap between families with different income levels, making it urgent to ensure students from various economic backgrounds can all benefit from LLMs equally. 

\section{Discussion and Conclusion}

The survey shows the use of LLMs in secondary education is widespread and is expected to grow further with the advent of more capable models and increased awareness. It also shows that resource-rich students are more likely to take advantage of LLMs with their studies, potentially causing underserved students with limited learning resources to fall further behind. We therefore propose a few ideas to address the feedback and to make LLMs more useful to students of all backgrounds.

\textbf{AI models should be fine-tuned for student usage.} Rather than restricting the use of LLMs in education, educators and AI developers should collaborate to build fine-tuned models trained on textbooks and teacher inputs \cite{bassner}. Such models will hallucinate less and be more relevant to topics at hand. Firstly, the LLMs must integrate a diverse range of educational resources to allow the model to generate well-rounded and informed responses. Secondly, they should be trained with teacher inputs that provide additional context on topics to encourage critical thinking by prompting students to explore topics more deeply. Lastly, safeguards must be implemented to prevent the generation of inappropriate or harmful content and protect privacy.  

\textbf{AI tutors can personalize learning.} An AI tutor will make students' educational experiences more accessible, comprehensive and engaging due to its easy access, vast knowledge and context-aware conversation capabilities \cite{thomas}. This is particularly beneficial to students in remote areas, those with disabilities, or learners who need additional help outside of traditional classrooms. AI tutors can also provide content in multiple languages and formats, catering to diverse learning needs and making education more inclusive. They can be made more useful by creating customized learning paths that adapt to a student's learning habits, providing immediate feedback and tailored quizzes, and aligning with the curriculum to address specific learning objects of various subjects and grades. 

\textbf{AI classrooms promote learning equality.} LLMs provide an opportunity for AI developers and teachers to work together to level the learning field for students from different demographics. We propose the idea of “AI classrooms:” the “AI teachers” in the virtual classrooms are trained on teaching videos and notes of experienced teachers. Using the multi-modal capabilities of LLMs, the “AI teachers” can "teach" and interact with the students in real-time. They will also proactively ask students questions and quiz them to provoke critical thinking skills. These multi-modal capabilities can also be personalized for differing learning paces. With government and industry support, the “AI classrooms” can be made free for those under-resourced districts as a supplement to students' education in school. 

\textbf{Limitations of the survey.} Respondents all had access to the internet in order to take the forms, introducing accessibility bias, which could lead to a higher LLM usage percentage than the national truth. The study provides educators, policymakers, and developers with information regarding LLMs in education but may not encompass all viewpoints due to the focus on students. 

{
\small
\bibliographystyle{plain}
\bibliography{custom}
}

\end{document}